\documentclass[3p,twocolumn,times]{elsarticle}

\usepackage{ecrc}

\volume{00}

\firstpage{1}

\journalname{Nuclear Physics B Proceedings Supplement}

\runauth{}

\jid{nuphbp}

\jnltitlelogo{Nuclear Physics B Proceedings Supplement}

\usepackage{amssymb}

\usepackage[figuresright]{rotating}

\newcommand{\lsi}{LS~I~+61$^\circ$~303}

\begin{document}

\begin{frontmatter}



\dochead{}

\title{Search for a neutrino flux from \lsi\ based on a time
  dependent model with IceCube}


\author[me]{L.~Demir\"ors}
\address[me]{Laboratoire de Physique des Hautes Energies, EPFL, CH-1015 Lausanne, Switzerland}

\author{for the IceCube collaboration}
\begin{abstract}
  \noindent We present a search based on a time dependent neutrino flux
  prediction from the X--ray binary \lsi. Results from data taken
  with the 22 and 40 strings of IceCube are compatible with background
  fluctuations.
\end{abstract}




\end{frontmatter}


\lsi\ is one of the most intensely studied galactic objects
(c.f.~\cite{Massi:2005fm}). Belonging to a group of
objects termed gamma--ray loud X--ray binaries (GRLB), it consists of a
massive central Be star orbited by a compact companion of unknown
nature (neutron star or black hole). It shows variable emission from
radio \cite{gregory-radioflux2} to X--rays, e.g.~\cite{Abdo:2009pw},
and has been detected at TeV energies
\cite{magic,Acciari:2009gg}. As yet, it is unclear if the observed
broadband emission is caused mainly by hadronic or leptonic
interactions. The observation of a neutrino flux would prove hadronic
acceleration at the source. 

In the model considered here \cite{lsi,Chernyakova:2009hp} neutrinos
originate from interactions of high energy (HE) protons from the compact
companion with the dense disk around the central star. For them to be
observable, they have to be produced in a cone
aligned with the line of sight. Since the HE protons will
only be weakly deflected by the system's magnetic field, highest
fluxes are expected when the companion is eclipsed by the stellar disk. During that phase the accompanying
$\gamma$--ray emission from secondary e$^{\pm}$ and $\pi^0$ decay is
strongly suppressed due to the high density of soft photons and matter
in the stellar disk. Therefore, the predicted neutrino flux does not
positively correlate with the observed TeV photon flux.

Two parameters determine the width of the phase where neutrino
emission can be observed: the inclination of the system's ecliptic with respect to
the observer, and the disk size. Four configurations were chosen to cover
scenarios where the companion is either a neutron star or black
hole. To account for the unknown acceleration mechanisms at the
source, two simple power law neutrino spectra were chosen with
spectral indices of $\Gamma = 1.5$ or $2$, respectively.


The IceCube observatory is a one cubic kilometer neutrino telescope
scheduled to be fully instrumented in 2011. The eight model
predictions from above were tested with 22--string data
taken in 2006--2007 \cite{ic22-ptsrc}. Assuming their validity, the sensitivity
of this search was increased in comparison to generic searches by
down-scaling the background expectation. Event candidates were
selected according to their distance to the source location, their
estimated energy and their time in the \lsi\ period. These search
parameters were optimized to give the best sensitivity for a $5\sigma$
discovery in $90\%$ of trials. The best model had four event
candidates within the search parameters. The preliminary post-trial p-value was
5.61\%. In a preliminary analysis on data taken with the
40--string configuration (2007--2008), no signal was seen.

This work is supported by the Swiss National Research Foundation
(grant PP002--114800)\\[1mm]




\noindent\textbf{References}
\bibliographystyle{elsarticle-num}
\bibliography{nuphbp-template}

\begin{thebibliography}{1}
\expandafter\ifx\csname url\endcsname\relax
  \def\url#1{\texttt{#1}}\fi
\expandafter\ifx\csname urlprefix\endcsname\relax\def\urlprefix{URL }\fi
\expandafter\ifx\csname href\endcsname\relax
  \def\href#1#2{#2} \def\path#1{#1}\fi

\bibitem{Massi:2005fm}
M.~Massi, {Introduction to Astrophysics of Microquasars} (2005).

\bibitem{gregory-radioflux2}
P.~C. Gregory, Astrophys. J. 575 (2002) 472.

\bibitem{Abdo:2009pw}
A.~A. Abdo, et~al., Astrophys. J. 701 (2009) L123--L128.

\bibitem{magic}
J.~Albert, et~al., Astrophys. J. 693 (2009) 303--310.

\bibitem{Acciari:2009gg}
V.~A. Acciari, et~al., Astrophys. J. 700~(2) (2009) 1034.

\bibitem{lsi}
A.~Neronov, M.~Ribordy, Phys. Rev. D 79~(4) (2009) 043013.

\bibitem{Chernyakova:2009hp}
M.~Chernyakova, A.~Neronov, M.~Ribordy, arXiv:0912.3821v1.

\bibitem{ic22-ptsrc}
R.~Abbasi, et~al., Astrophys. J. 701 (2009) L47--L51.

\end{thebibliography}







\end{document}